\documentclass[numberedheadings]{aip-cp} 

\usepackage[numbers,sort&compress]{natbib}
\usepackage{rotating}
\usepackage{graphicx}
\usepackage{color}
\usepackage{subfigure}

\begin{document}

\title{Transverse Magnetic Susceptibility of a Frustrated Spin-$\frac{1}{2}$  $J_{1}$--$J_{2}$--$J_{1}^{\perp}$ Heisenberg Antiferromagnet on a Bilayer Honeycomb Lattice}

\author[aff1,aff2]{P.H.Y. Li}
\eaddress{peggyhyli@gmail.com}
\author[aff1,aff2]{ and \underline{R. F. Bishop}}
\eaddress{raymond.bishop@manchester.ac.uk}

\affil[aff1]{School of Physics and Astronomy, The University of Manchester, Schuster Building, Manchester, M13 9PL, United Kingdom}
\affil[aff2]{School of Physics and Astronomy, University of Minnesota, 116 Church Street SE, Minneapolis, Minnesota 55455, USA}

\maketitle

\begin{abstract}
  We use the coupled cluster method (CCM) to study a frustrated
  spin-$\frac{1}{2}$ $J_{1}$--$J_{2}$--$J_{1}^{\perp}$ Heisenberg
  antiferromagnet on a bilayer honeycomb lattice with $AA$ stacking.
  Both nearest-neighbor (NN) and frustrating next-nearest-neighbor
  antiferromagnetic (AFM) exchange interactions are present in each
  layer, with respective exchange coupling constants $J_{1}>0$ and
  $J_{2} \equiv \kappa J_{1} > 0$.  The two layers are coupled with NN
  AFM exchanges with coupling strength
  $J_{1}^{\perp}\equiv \delta J_{1}>0$.  We calculate to high orders
  of approximation within the CCM the zero-field transverse magnetic
  susceptibility $\chi$ in the N\'{e}el phase.  We thus obtain an
  accurate estimate of the full boundary of the N\'{e}el phase in the
  $\kappa\delta$ plane for the zero-temperature quantum phase diagram.
  We demonstrate explicitly that the phase boundary derived from
  $\chi$ is fully consistent with that obtained from the vanishing of
  the N\'{e}el magnetic order parameter.  We thus conclude that at all
  points along the N\'{e}el phase boundary quasiclassical magnetic
  order gives way to a nonclassical paramagnetic phase with a nonzero
  energy gap.  The N\'{e}el phase boundary exhibits a marked reentrant
  behavior, which we discuss in detail.
\end{abstract}

\section{INTRODUCTION}

\label{introd_sec}
The study of the zero-temperature $(T=0)$ quantum phase transitions
(QPTs) that can occur in strongly-interacting quantum many-body
systems has become a field of huge current interest
\cite{Sachdev:2008_QPT,Sachdev:2011_QPT} in condensed matter physics.
Quantum spin-lattice models have become a particularly interesting
test-bed in this context since they exhibit such a wide variety of
phases at $T=0$, ranging from (partially) ordered quasiclassical
states with magnetic long-range order (LRO) to such nonclassical
states as those with valence-bond crystalline (VBC) order and various
quantum spin liquid (QSL) phases.  All such quantum magnets basically
comprise an extended regular periodic lattice in $d$ dimensions of a
specified form, on each of the $(N \rightarrow \infty)$ sites of which
is situated a magnetic ion with spin quantum number $s$.

A particularly important class of such spin-lattice models comprises
those in which the spins interact only via a pairwise term in the
Hamiltonian of the isotropic Heisenberg form
$J_{ij}\mathbf{s}_{i}\cdot \mathbf{s}_{j}$, between spins at lattice
sites $i$ and $j$.  If we restrict the exchange couplings $J_{ij}$ to
be only between nearest-neighbor (NN) pairs of sites on the lattice,
and further restrict all NN pairs to be magnetically equivalent, so
that each pair feels the same exchange coupling $J_{1}$, then the only
relevant parameter (for a given lattice type and for specified values
of $d$ and $s$) is the sign of $J_{1}$, since $|J_{1}|$ simply sets
the overall energy scale.  For $J_{1}<0$, the energy of the system is
minimized when all of the spins align, and the system takes perfect
ferromagnetic LRO at $T=0$.  This is true both classically (i.e., when
$s \rightarrow \infty$) and at the quantum level (i.e., for finite
values of $s$), since the (classical) state with all the spins aligned
in some direction is also then an eigenstate of the quantum
Hamiltonian.

Conversely, when $J_{1}>0$, the isotropic Heisenberg interaction
prefers to anti-align spins on NN sites.  For the simplest example of
a bipartite lattice, which contains no geometric frustration, this
then leads classically (i.e., for $s \rightarrow \infty$) to the
N\'{e}el state, with perfect antiferromagnetic (AFM) LRO, being the
ground-state (GS) phase of the system at $T=0$.  The quantum situation
(i.e., for $s$ finite) is more subtle, since such a N\'{e}el state
is now {\it not} an eigenstate of the Heisenberg antiferromagnet
(HAF), even in the case of NN interactions only.  The interesting
question then arises as to whether such unfrustrated HAFs exhibit
magnetic N\'{e}el LRO at all.  It is clear that the effect of quantum
fluctuations in all such cases will be to reduce the order parameter
$M$ (viz., the sublattice magnetization or average local on-site
magnetization) from its classical value of $s$.  The real question is
whether, for a given lattice and given values of $d$ and $s$, $M$ is
reduced to zero or takes a nonzero value in the range $0 < M < s$.
Now the Mermin-Wagner theorem \cite{Mermin:1966} excludes the breaking
of any continuous symmetry, and hence of any form of magnetic LRO, in
any spin-lattice problem with finite $s$ in which all the interactions
are of the isotropic Heisenberg from, both for $d=1$ (even at $T=0$)
and for $d=2$ except precisely at $T=0$.

For this reason, spin-lattice models with $d=2$ and at $T=0$, now play
a very special role in the study of QPTs.  In this context perhaps the
simpest class of lattices is that comprising the Archimedean lattices, which
are defined for $d=2$ to have all sites equivalent to one another and
to be composed only of regular polygons.  They are defined uniquely by
specifying the ordered sequence of polygons that surrounds each
(equivalent) vertex.  Of the eleven Archimedean lattices for $d=2$,
seven include triangles and are hence geometrically frustrated for the
formation of N\'{e}el AFM states.  Examples include the triangle
$(3^{6})$, kagome $(3,6,3,6)$, and star $(3,12^{2})$ lattices.  The
remaining four Archimedean lattices have all of the polygons
even-sided and are hence bipartite.  Two of these, namely the square
$(4^{4})$ and honeycomb $(6^{3})$ lattices, also have all of the edges
(or NN bonds in our magnetic spin-lattice language) equivalent.  The
other two, namely the CaVO $(4,8^{2}$) and SHD $(4,6,12)$ lattices,
contain nonequivalent edges (i.e., with two different types of NN
bonds).

By now it is well established that all four unfrustrated HAFs (with NN
interactions) on the bipartite Archimedean lattices have N\'{e}el
magnetic LRO
\cite{Richter:2004_triang_ED,DJJFarnell:2014_archimedeanLatt} albeit
with values of $M$ significantly reduced by quantum fluctuations from
the classical value, defined to be $s$.  We expect, {\it a priori},
the role of quantum fluctuations (for $d=2$) to be greater for smaller
values of both $s$ and the lattice coordination number $z$ (i.e., the
number of NN sites to a given site), other things being equal.  The
four bipartite Archimedean lattices have $z=3$ for the honeycomb
lattice and $z=4$ for each of the square, CaVO, and SHD lattices.  In
the case when $s=\frac{1}{2}$, the HAF models (with NN interactions
only) have $M \approx 0.31$ for the square lattice and
$M \approx 0.27$ for the honeycomb lattice
\cite{DJJFarnell:2014_archimedeanLatt}, in line with our expectations
for the relative effect of quantum fluctuations for different values
of $z$.  Interestingly, however, the reduction in N\'{e}el order is
even stronger for the two bipartite Archimedean lattices with
nonequivalent NN edges (i.e, with two different sorts of NN bonds),
which have values $M \approx 0.22$ for the CaVO lattice and
$M \approx 0.18$ for the SHD lattice
\cite{DJJFarnell:2014_archimedeanLatt}.  This is almost certainly an
indication of an emergent instability for these two HAFs against the
formation of a paramagnetic VBC state (and see, eg., Ref.\
\cite{Troyer:1996_cavo_qmc}), as the relative strength of the two
different types of bonds on the inequivalent edges would be varied,
for example.

Of all the bipartite lattices with all sites and edges equivalent to
one another, the honeycomb lattice has the lowest coordination number,
$z=3$, and hence the greatest expected effect of quantum fluctuations.
Similarly, we expect the largest deviations from classical behavior
for spins with $s=\frac{1}{2}$.  Hence, it is natural for
spin-$\frac{1}{2}$ models on the honeycomb lattice to occupy a special
role in the study of QPTs.  However, since the spin-$\frac{1}{2}$ HAF
on the honeycomb monolayer, with NN interactions only, has N\'{e}el
LRO, this order can only be destroyed by the addition of competing
interactions.  There are two relatively simple ways to do this that
spring immediately to mind.  One is to include isotropic AFM
Heisenberg interactions between next-nearest-neighbor pairs (NNN) of spins,
all with equal exchange coupling constant,
$J_{2}\equiv\kappa J_{1} > 0$.  The resulting so-called
$J_{1}$--$J_{2}$ model on the honeycomb lattice has been intensively
studied, using a wide variety of theoretical techniques \cite{Rastelli:1979_honey,Mattsson:1994_honey,Fouet:2001_honey,Mulder:2010_honey,Ganesh:2011_honey,Ganesh:2011_honey_errata,Clark:2011_honey,Reuther:2011_honey,Albuquerque:2011_honey,Mosadeq:2011_honey,Oitmaa:2011_honey,Mezzacapo:2012_honey,Li:2012_honey_full,Bishop:2012_honeyJ1-J2,RFB:2013_hcomb_SDVBC,Zhang:2013_honey,Ganesh:2013_honey_J1J2mod-XXX,Zhu:2013_honey_J1J2mod-XXZ,Gong:2013_J1J2mod-XXX,Yu:2014_honey_J1J2mod}.
Clearly, the $J_{1}$ and $J_{2}$ bonds act to frustrate one another.

A second method to include competing bonds, this time {\it without}
frustration, is to consider a honeycomb bilayer in $AA$ stacking
(i.e., with the two layers stacked so that every site on one layer is
immediately above its counterpart on the other layer), and now include
an interlayer NN interaction of the same isotropic AFM Heisenberg
type, with all interlayer NN bonds having the same exchange coupling
strength $J_{1}^{\perp} \equiv \delta J_{1} > 0$.  While the
$J_{1}^{\perp}$ bonds do not frustrate the $J_{1}$ bonds, since both
act to promote anti-aligned NN pairs, they are nevertheless in
competition for the quantum models (i.e., with finite values of $s$).
This is because the $J_{1}^{\perp}$ bonds acting alone promote the
formation of interlayer spin-singet dimers, and hence there is now
competition between a phase with N\'{e}el magnetic LRO and a
nonclassical paramagnetic interlayer-dimer VBC (IDVBC) phase.  The
effect of including the $J_{1}^{\perp}$ bonds is thus rather similar
to that in the monolayer CaVO or SHD lattices in the case that the two
topologically inequivalent types of NN bonds are allowed to have
different strengths.

This N\'{e}el to dimer transition has been studied, using an exact
stochastic series-expansion quantum Monte Carlo (QMC) technique
\cite{Ganesh:2011_honey_bilayer_PRB84}, for spin-$s$ HAFs on the
bilayer honeycomb lattice in the so-called $J_{1}$--$J_{1}^{\perp}$
model.  Nevertheless, it is clearly of considerable interest to examine
a model in which both types of competition to destroy N\'{e}el order
act together.  For that reason we study here the so-called
$J_{1}$--$J_{2}$--$J_{1}^{\perp}$ model on a honeycomb bilayer, for
the case of spins with $s=\frac{1}{2}$.  This model was studied
recently \cite{Zhang:2014_honey_bilayer,Arlego:2014_honey_bilayer}
using Schwinger-boson mean-field theory.  Although the results were
compared with those from exactly diagonalizing a relatively small
(24-site) cluster and from a dimer-series expansion for the
spin-triplet energy gap carried out only to low (viz., fourth) orders,
such mean-field results cannot be regarded as fully converged and,
hence, as fully reliable.

For that reason, in a recent paper
\cite{Bishop:2017_honeycomb_bilayer_J1J2J1perp} we studied the model
using a high-order implementation of a fully microscopic quantum
many-body theory approach, namely the coupled cluster method (CCM)
\cite{Coester:1958_ccm,Coester:1960_ccm,Cizek:1966_ccm,Kummel:1978_ccm,Bishop:1978_ccm,Bishop:1982_ccm,Arponen:1983_ccm,Bishop:1987_ccm,Arponen:1987_ccm,Arponen:1987_ccm_2,Bartlett:1989_ccm,Arponen:1991_ccm,Bishop:1991_TheorChimActa_QMBT,Bishop:1998_QMBT_coll,Zeng:1998_SqLatt_TrianLatt,Fa:2004_QM-coll},
which has very successfully been applied in the past to a wide variety
of systems in quantum magnetism, including corresponding monolayer
honeycomb-lattice models
\cite{Li:2012_honey_full,Bishop:2012_honeyJ1-J2,RFB:2013_hcomb_SDVBC,DJJF:2011_honeycomb,PHYLi:2012_honeycomb_J1neg,Bishop:2012_honey_circle-phase,Bishop:2014_honey_XY,Li:2014_honey_XXZ,Bishop:2014_honey_XXZ_nmp14,Bishop:2015_honey_low-E-param,Bishop:2016_honey_grtSpins}
to that considered here.  In our earlier work
\cite{Bishop:2017_honeycomb_bilayer_J1J2J1perp} on the
$J_{1}$--$J_{2}$--$J_{1}^{\perp}$ model on the honeycomb bilayer
lattice we gave results for the GS energy per spin, the N\'{e}el
magnetic order parameter $M$, and the triplet spin gap $\Delta$, all
as functions of both $\kappa$ and $\delta$.  Information on $M$ and
$\Delta$ was used to construct the full phase boundary of the N\'{e}el
phase in the $\kappa\delta$ plane.  In the present work our aim is to
augment the earlier results by calculations of the zero-field
transverse (uniform) magnetic susceptibility $\chi$ in the same
$\kappa\delta$ window.  A particular advantage of studying $\chi$ is
that it can carry information about both the melting of the N\'{e}el
phase and whether or not a gapped phase emerges at the corresponding
QCP \cite{Mila:2000_M-Xcpty_spinGap,Bernu:2015_M-Xcpty_spinGap}.  We
thereby provide independent confirmation of our previous results on
the N\'{e}el phase boundary, together with a non-biased indication of
their accuracy.

The plan of the remainder of the paper is as follows.  The model
itself is first discussed in more detail in Sec.\ \ref{model_sec},
where we also describe what is known for the monolayer case
$(\delta=0)$ of our model.  In Sec.\ \ref{ccm_section} we then briefly
describe the most important features of the CCM, before discussing our
results in Sec.\ \ref{results_section}.  We conclude with a discussion
and summary in Sec.\ \ref{summary_section}.

\section{THE MODEL}
\label{model_sec}
The $J_{1}$--$J_{2}$--$J_{1}^{\perp}$ model on the
honeycomb bilayer lattice is specified by Hamiltonian,
\begin{eqnarray}
H=&&J_{1}\sum_{{\langle i,j \rangle},\alpha} \mathbf{s}_{i,\alpha}\cdot\mathbf{s}_{j,\alpha} + 
J_{2}\sum_{{\langle\langle i,k \rangle\rangle},\alpha} \mathbf{s}_{i,\alpha}\cdot\mathbf{s}_{k,\alpha} +  J_{1}^{\perp}\sum_{i} \mathbf{s}_{i,A}\cdot\mathbf{s}_{i,B}\,,
\label{H_eq}
\end{eqnarray}
where the two layers are labeled by the index $\alpha=A,B$.  Each site $i$
of both layers of the honeycomb lattice carries a spin-$s$ particle,
which is described in terms of the usual SU(2) spin operators
${\bf
  s}_{i,\alpha}\equiv(s^{x}_{i,\alpha},s^{y}_{i,\alpha},s^{z}_{i,\alpha})$,
with ${\bf s}^{2}_{i,\alpha} = s(s+1)$.  For the present work we
restrict attention to the case $s=\frac{1}{2}$.  In Eq.\ (\ref{H_eq})
the sums over $\langle i,j \rangle$ and
$\langle \langle i,k \rangle \rangle$ run respectively over all NN and
NNN intralayer pairs, counting each bond once only in each sum.
Similarly, the last sum in Eq.\ (\ref{H_eq}) includes all NN
interlayer bonds.  We will be interested here in the case when all
three bonds are AFM in nature (i.e., $J_{1}>0$, $J_{2}>0$,
$J_{1}^{\perp}>0$).  The two relevant parameters of the model are thus
$J_{2}/J_{1} \equiv \kappa$ and $J_{1}^{\perp}/J_{1}\equiv\delta$,
since we may regard $J_{1}$ as a multiplicative constant in the
Hamiltonian that simply sets the overall energy scale.

\begin{figure*}[!t]
\centerline{
\mbox{
\subfigure[]{\includegraphics[width=4.0cm]{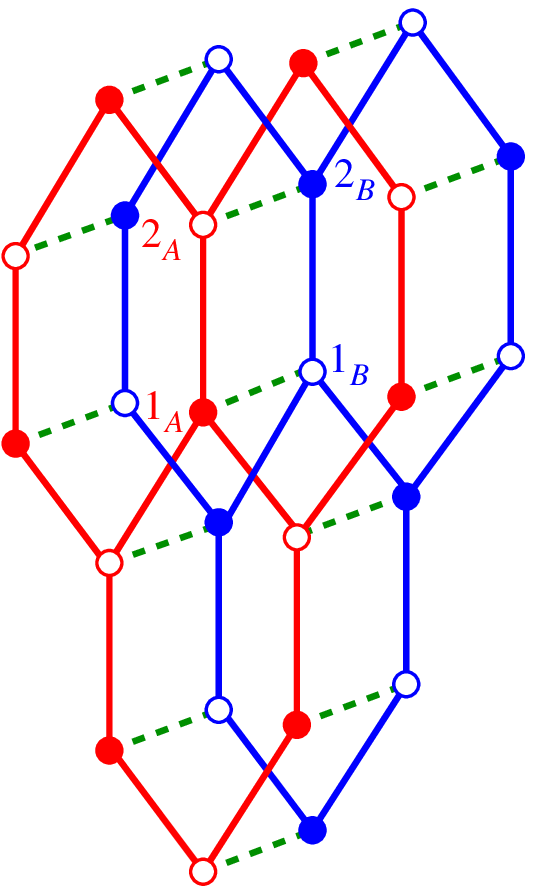}}
\hspace{2cm} \subfigure[]{\includegraphics[width=3.5cm]{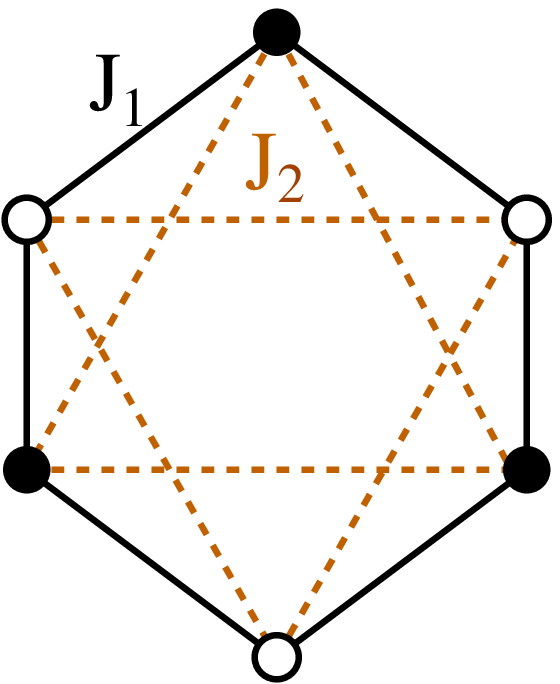}}
}
}
\caption{The $J_{1}$--$J_{2}$--$J_{1}^{\perp}$ model on the honeycomb bilayer
    lattice, showing (a) the two layers $A$ (red) and $B$ (blue), the nearest-neighbor bonds ($J_{1} = $ -----; $J_{1}^{\perp} = $ - - -), and the four sites ($1_{A}, 2_{A}, 1_{B}, 2_{B}$) of the unit cell; and (b) the intralayer bonds $J_{1} = $ -----; $J_{2} = $ - - -) on each layer.  Sites on the two monolayer triangular sublattices are shown by filled and empty circles respectively.}
\label{model_pattern}
\end{figure*}

The honeycomb monolayer lattice is non-Bravais.  Its unit cell
contains two sites, with two interlacing triangular sublattices 1 and
2, shown in Fig.\ \ref{model_pattern} by filled and empty circles.
The corresponding $AA$-stacked bilayer unit cell thus contains 4
sites, as shown explicitly in Fig.\ \ref{model_pattern}(a).  The three
types of AFM bonds are also shown in Fig.\ \ref{model_pattern}.

Let us first consider the classical limit ($s \rightarrow \infty$) of
the model.  For the honeycomb-lattice monolayer (i.e., for $\delta=0$)
N\'{e}el AFM order (i.e., where all spins on lattice sites denoted by
filled circles in Fig.\ \ref{model_pattern} point in a given,
arbitrary, direction, and those on the sites denoted by empty circles
point in the opposite direction) persists for all values
$\kappa \leq \frac{1}{6}$ of the intralayer frustration parameter
\cite{Rastelli:1979_honey,Fouet:2001_honey}.  At this QCP a phase
transition to a state with spiral order occurs.  Indeed, the GS phase
for $\kappa > \frac{1}{6}$ has a spiral wave vector that can point in
an arbitrary direction.  As a consequence, there now exists an
infinite classical one-parameter family of states, all degenerate in
energy.  Spin-wave fluctuations have been shown to lift this
accidental degeneracy by favouring particular wave vectors
\cite{Mulder:2010_honey}.  This mechanism has hence become known as
spiral order by disorder.

By comparison with the classical case one would expect that in the
quantum case the critical value of $\kappa$ at which N\'{e}el order
melts will be {\it larger} than the classical value of $\frac{1}{6}$,
since quantum fluctuations as a general rule tend to favor collinear
phases over spiral phases.  There is by now a wide consensus that this
expectation is fulfilled by the present model, with a large number of
calculations for the $s=\frac{1}{2}$ $J_{1}$--$J_{2}$
honeycomb-lattice monolayer giving a critical value of $\kappa$ for
the vanishing of N\'{e}el order in the approximate range
$0.17$--$0.22$
\cite{Albuquerque:2011_honey,Mosadeq:2011_honey,Mezzacapo:2012_honey,Bishop:2012_honeyJ1-J2,RFB:2013_hcomb_SDVBC,Zhang:2013_honey,Ganesh:2013_honey_J1J2mod-XXX,Zhu:2013_honey_J1J2mod-XXZ,Gong:2013_J1J2mod-XXX,Yu:2014_honey_J1J2mod}.
There is also broad agreement, including from calculations using the
CCM \cite{Bishop:2012_honeyJ1-J2,RFB:2013_hcomb_SDVBC} that we employ
here, that spiral order is absent for the spin-$\frac{1}{2}$ case over
the entire range $0 \leq \kappa \leq 1$ of the frustration parameter.

If we now turn our attention to the $J_{1}$--$J_{2}$--$J_{1}^{\perp}$
model on the bilayer honeycomb lattice, at the classical level (i.e.,
when $s \rightarrow \infty$) the introduction of the interlayer NN
coupling $J_{1}^{\perp}$ is essentially trivial.  Thus, the classical
(N\'{e}el and spiral) phases are totally unaffected, since the
$J_{1}^{\perp}$ coupling introduces no extra frustration.  The NN
interlayer pairs simply anti-align (for the case when
$J_{1}^{\perp}>0$, as considered here), and the order in each layer
remains unchanged.  However, for the quantum versions of the model
(i.e., for finite values of $s$) the situation differs significantly,
since for large enough values of the parameter
$\delta \equiv J_{1}^{\perp}/J_{1}$, for a fixed value of the
frustration parameter $\kappa \equiv J_{2}/J_{1}$, we clearly expect
the GS phase to be an IDVBC phase.  As we mentioned previously in
Sec.\ \ref{introd_sec}, the CCM has been employed very recently
\cite{Bishop:2017_honeycomb_bilayer_J1J2J1perp} to study the GS phase
diagram of the spin-$\frac{1}{2}$ $J_{1}$--$J_{2}$--$J_{1}^{\perp}$
model on a bilayer honeycomb lattice in the $\kappa\delta$ plane,
particularly to find the region of stability of the N\'{e}el phase.
In that earlier work it was found that the best estimate for the
N\'{e}el phase boundary came from the vanishing of the calculated
N\'{e}el order parameter $M$.  Calculations of the triplet spin gap
$\Delta$ were also performed.  While these calculations corroborated
the estimates from $M$, in practice they were less accurate.  Our aim
now is to calculate for the same system the zero-field transverse
magnetic susceptibility $\chi$, in order to find further corroboration
of the earlier results.

The N\'{e}el state that we envisage is clearly now one in which the
spins on all of the sites in Fig.\ \ref{model_pattern}(a) shown by
filled circles point in a given (arbitrarily chosen) direction and
those on the sites shown by empty circles point in the opposite
direction.  Let us now apply an external magnetic field of strength
$h$ in a direction perpendicular to the N\'{e}el alignment direction
(and we choose units such that the gyromagnetic ratio
$g\mu_{B}/\hbar=1$).  The spins will thus cant at an angle $\alpha$
with respect to their zero-field configurations, and $\alpha$ may be
found by minimizing the energy $E = E(h)$ in the presence of the
field.  The (uniform) transverse magnetic susceptibility, $\chi(h)$,
is then defined, as usual, to be
\begin{equation}
\chi({h})=-\frac{1}{N}\frac{{\rm d}^{2}E}{{\rm d}h^{2}}\,.    \label{chi_eq}
\end{equation}
Its zero-field limit, $\chi \equiv \chi(0)$, in which we are
interested here, is one of the parameters of the effective magnon
field theory that fully describes the low-energy behavior of the
system.  For the classical ($s \rightarrow \infty$) version of our
model it is easy to calculate its value in the N\'{e}el phase to be,
\begin{equation}
\chi_{{\rm cl}}^{{\rm N\acute{e}el}}=\frac{1}{2J_{1}(3+\delta)}\,,  \label{chi_eq_neel}
\end{equation}
independent of the frustration parameter $\kappa$.

\section{THE COUPLED CLUSTER METHOD}
\label{ccm_section}
The CCM
\cite{Coester:1958_ccm,Coester:1960_ccm,Cizek:1966_ccm,Kummel:1978_ccm,Bishop:1978_ccm,Bishop:1982_ccm,Arponen:1983_ccm,Bishop:1987_ccm,Arponen:1987_ccm,Arponen:1987_ccm_2,Bartlett:1989_ccm,Arponen:1991_ccm,Bishop:1991_TheorChimActa_QMBT,Bishop:1998_QMBT_coll,Zeng:1998_SqLatt_TrianLatt,Fa:2004_QM-coll}
provides one of the most accurate and most adaptable {\it ab inito}
techniques of modern quantum many-body theory.  It is both
size-consistent and size-extensive at every level of approximation,
thereby ensuring that the method can be implemented in the
infinite-lattice ($N \rightarrow \infty$) limit from the outset.
Thus, no finite-size scaling is ever needed.  Since this is often a
large source of errors in many competing methods, it is a considerable
advantage of using the CCM.  Further advantages are that the very
important Hellmann-Feynman theorem is also preserved at every level of
approximation, together with the Goldstone linked-cluster theorem.
These ensure that the method provides accurate, robust, and
self-consistent results for a variety of calculated physical
parameters for any specific system.  The CCM can furthermore nowadays
be implemented computationally to high orders of approximation in
well-studied and well-understood truncation hierarchies that become
exact as some specified parameter that describes the order of the
approximation approaches infinity.  The {\it only} approximation ever
made in the CCM is thus to extrapolate the sequences of calculated
approximants for any specified parameter of the system in which we are
interested.  By now there are well-studied and well-understood
extrapolation schemes available for a wide variety of physical
parameters.

For present purposes we will very briefly review here only some of the
principal and most pertinent features of the CCM as it is applied to
quantum spin-lattice models, and refer the interested reader to the by now
very extensive literature (and see, e.g., Refs.\
\cite{Zeng:1998_SqLatt_TrianLatt,Fa:2004_QM-coll} in particular) for
full details.  In order to utilize the CCM in practice, the first step
is always to choose a suitable model (or reference) state
$|\Phi\rangle$, which acts as a generalized vacuum state, and with
respect to which the quantum correlations present in the exact GS wave
function $|\Psi\rangle$ can then later be incorporated in a systematic
way.  For spin-lattice systems all (quasi)classical states with
perfect magnetic LRO provide suitable such model states.  Here we will
use both the N\'{e}el state and its canted equivalent in the presence of an
external transverse magnetic field as our CCM model states.  For later
purposes it is extremely convenient to be able to consider all lattice
spins as being fully equivalent to each other in {\it every} model
state.  In particular, this will then allow us to use a universal
computational technique that is suitable for any spin-lattice model
(at least initially for the phases with quasiclassical order) \cite{ccm_code}.  An obvious way to do this is clearly to make a
passive rotation of each spin separately (in any such classical model
state) so that, in their own set of local spin-coordinate frames, they
all point in the same direction, say downwards (i.e., along the local
negative $z_{s}$ axis).  Every model state will thus take the
universal from
$|\Phi\rangle =
|\downarrow\downarrow\downarrow\cdots\downarrow\rangle$ in its own set
of local frames.  Evidently we still need to rewrite the Hamiltonian of
the system as appropriate in the specific choice of local
spin-coordinate frames.

The exact GS wave function $|\Psi\rangle$,  where $H|\Psi\rangle=E|\Psi\rangle$, is now expressed within the CCM in the exponentiated form,
\begin{equation}
|\Psi\rangle={\rm e}^{S}|\Phi\rangle\,; \quad S=\sum_{I\neq 0}{\cal S}_{I}C^{+}_{I}\,,  \label{ket_eigen_eq}
\end{equation}
that is distinctive for the method.  The set-index $I$ represents a multispin configuration, such that the set of states $\{C_{I}^{+}|\Phi\rangle\}$ completely spans the ket-state Hilbert space.  We choose 
$C^{+}_{0}\equiv 1$ to be the identity operator.  Clearly, with the CCM model state chosen as above to be in the universal form $|\Phi\rangle = |\downarrow\downarrow\downarrow\cdots\downarrow\rangle$ in an appropriate choice of locally rotated spin-coordinate frames, the operator 
$C^{+}_{I}$ now also takes the universal form of a
product of single-spin raising operators, $s^{+}_{k}
\equiv s^{x}_{k}+is^{y}_{k}$.   The set index $I$ now is expressed as a set of lattice site indices,
\begin{equation}
I \equiv \{k_{1},k_{2},\cdots , k_{n};\; \quad n=1,2,\cdots , 2sN\}\,,
\end{equation}
in which no given site index $k_{i}$ may appear more than $2s$ times
(for spins of general spin quantum number $s$).  The operator
$C^{+}_{I}$ thereby creates a multispin configuration cluster,
\begin{equation}
C^{+}_{I} \equiv s^{+}_{k_{1}}s^{+}_{k_{2}}\cdots s^{+}_{k_{n}};\; \quad
n=1,2,\cdots , 2sN\,.
\end{equation}
The model state $|\Phi\rangle$ and the complete set of mutually
commuting multispin creation operators $\{C_{I}^{+}\}$,
\begin{equation}
[C^{+}_{I},C^{+}_{J}]=0\,, \quad \forall I,J\,,   \label{commute-relation-create-destruct-oper}
\end{equation}
are hence chosen so that $|\Phi\rangle$ is a fiducial vector (or
generalized vacuum state) with respect to the set $\{C_{I}^{+}\}$, and
hence so that the latter obey the conditions,
\begin{equation}
\langle\Phi|C^{+}_{I} = 0 = C^{-}_{I}|\Phi\rangle\,, \quad \forall I
\neq 0\,,  \label{creat-destruct-operators-relationship}
\end{equation}
where $C^{-}_{I} \equiv (C^{+}_{I})^{\dagger}$ is the corresponding
multispin destruction operators.  The states
$\{C_{I}^{+}|\Phi\rangle\}$ are also usefully orthonormalized, so that
they obey the relations
\begin{equation}
\langle\Phi|C_{I}^{-}C_{J}^{+}|\Phi\rangle=\delta_{I,J}\,, \quad \forall I,J \neq 0\,,
\end{equation}
with $\delta_{I,J}$ defined as a generalized Kronecker symbol.

We note that the model state $|\Phi\rangle$ is (always) chosen to be
normalized, $\langle\Phi|\Phi\rangle=1$, and the CCM parametrization
of Eq.\ (\ref{ket_eigen_eq}) automatically ensures that the exact GS
energy eigenket $|\Psi\rangle$ obeys the intermediate normalization
condition, $\langle\Phi|\Psi\rangle=1$, due to Eq.\
(\ref{creat-destruct-operators-relationship}).  In general, of course,
$\langle\Psi|\Psi\rangle\neq 1$.  The corresponding GS energy eigenbra
$\langle\tilde{\Psi}|$, which obeys the Schr\"{o}dinger equation
$\langle\tilde{\Psi}|H=E\langle\tilde{\Psi}|$, takes the CCM
parametrization,
\begin{equation}
\langle\tilde{\Psi}|=\langle\Phi|\tilde{S}{\rm e}^{-S}\,, \quad \tilde{S}=1+\sum_{I\neq 0}\tilde{{\cal S}}_{I}C^{-}_{I}\,. \label{bra_eigen_eq}
\end{equation}
Equation (\ref{bra_eigen_eq}) ensures the automatic fulfillment of the
normalization condition $\langle\tilde{\Psi}|\Psi\rangle=1$.  While
Hermiticity clearly implies that the CCM correlation correlation operators
$S$ and $\tilde{S}$ are connected via the relation,
\begin{equation}
\langle\Phi|\tilde{S} = \frac{\langle\Phi|{\rm e}^{S^{\dagger}}{\rm e}^{S}}{\langle\Phi|{\rm e}^{S^{\dagger}}{\rm e}^{S}|\Phi\rangle}\,,  \label{correlation-opererators-relationship}
\end{equation}
a key feature of the CCM is that this constraint is not explicitly
imposed.  Instead the $c$-number parameters $\{{\tilde{\cal S}}_{I}\}$
are considered to be formally independent of their $\{{\cal S}_{I}\}$
counterparts.  Clearly, the Hermiticity constraint of Eq.\
(\ref{correlation-opererators-relationship}) will be exactly fulfilled
in the exact limit when all multispin clusters specified by the
complete set of indices $\{I\}$ are retained in the CCM expansions of
Eqs.\ (\ref{ket_eigen_eq}) and (\ref{bra_eigen_eq}).  However, in
practice, when approximations are made, as described below, to
restrict ourselves to some suitable subset of the indices $\{I\}$,
Hermiticity may only approximately be fulfilled.  Nevertheless, it is
very important to realize that this partial loss of exact Hermiticity
is always more than compensated in practice by the exact fulfillment
of the Hellmann-Feynman theorem at every level of approximation.

All GS physical quantities may now be expressed entirely in terms of
the CCM correlation coefficients
$\{{\cal S}_{I}, \tilde{{\cal S}}_{I}\}$.  For example, the GS
magnetic order parameter $M$, which is just the average local on-site
magnetization, may be expressed as
\begin{equation}
M = -\frac{1}{N}\langle\Phi|\tilde{S}\sum^{N}_{k=1}
  {\rm e}^{-S}s^{z}_{k}{\rm e}^{S}|\Phi\rangle\,.   \label{M_eq}
\end{equation}
where $s_{k}^{z}$ is expressed in the local (rotated) spin-coordinate
frames described above.  The parameters
$\{{\cal S}_{I}, \tilde{{\cal S}}_{I}\}$ are themselves now formally
obtained by minimization of the energy expectation functional,
\begin{equation}
\bar{H}=\bar{H}[{\cal S}_{I},{\tilde{\cal S}_{I}}]\equiv
\langle\Phi|\tilde{S}{\rm e}^{-S}H{\rm e}^{S}|\Phi\rangle\,,  \label{GS_E_xpect_funct}
\end{equation}
with respect to each of them, considered as independent variables.

Thus, firstly, using the explicit parametrization of Eq.\
(\ref{bra_eigen_eq}), extremization of $\bar{H}$ from Eq.\
(\ref{GS_E_xpect_funct}) with respect to the parameter
${\tilde{\cal S}}_{I}$, yields the relations
\begin{equation}
\langle\Phi|C^{-}_{I}{\rm e}^{-S}H{\rm e}^{S}|\Phi\rangle = 0\,, \quad \forall I \neq 0\,.  \label{ket_eq}
\end{equation}
Equation (\ref{ket_eq}) is simply a coupled set of nonlinear equations
for the creation coefficients $\{{\cal S}_{I}\}$, with as many
equations as there are unknown.  Secondly, using the explicit CCM
parametrization of Eq.\ (\ref{ket_eigen_eq}), extremization of
$\bar{H}$ from Eq. (\ref{GS_E_xpect_funct}) with respect to the
parameters ${\cal S}_{I}$, yields the respective relations
\begin{equation}
\langle\Phi|\tilde{S}{\rm e}^{-S}[H,C^{+}_{I}]{\rm e}^{S}|\Phi\rangle=0\,, \quad \forall I \neq 0\,.  \label{bra_eq}
\end{equation}
By making use of the simple relation $[S, C_{I}^{+}]=0$, which follows
trivially from Eqs.\ (\ref{ket_eigen_eq}) and
(\ref{commute-relation-create-destruct-oper}), Eq.\ (\ref{bra_eq}) may
readily be expressed in the equivalent form,
\begin{equation}
\langle\Phi|\tilde{S}({\rm e}^{-S}H{\rm e}^{S}-E)C^{+}_{I}|\Phi\rangle=0\,, \quad \forall I \neq 0\,,  \label{bra_eq_alternative}
\end{equation}
where we have re-expressed the GS ket-state Schr\"{o}dinger equation in the form,
\begin{equation}
{\rm e}^{-S}H{\rm e}^{S}|\Phi\rangle=E|\Phi\rangle\,,
\end{equation}
using Eq.\ (\ref{ket_eigen_eq}).  Equation (\ref{bra_eq_alternative})
is thus a set of generalized linear eigenvalue equations for the
destruction coefficients $\{{\tilde{\cal S}_{I}}\}$, with the
coefficients $\{{\cal S}_{I}\}$ as known input from first solving Eq.\
(\ref{ket_eq}), again with as many equations as unknowns.

We note that the exponentiated forms ${\rm e}^{\pm S}$, which are such
a characteristic and distinctive element of the CCM, always only enter
in the form of a similarity transform
${\rm e}^{-S}\Omega{\rm e}^{S}$ of some operator $\Omega$, where
$\Omega=H$ in Eqs.\ (\ref{ket_eq}) and (\ref{bra_eq_alternative}),
which are the equations to be solved for
$\{{\cal S}_{I},{\tilde{\cal S}}_{I}\}$, and $\Omega=s_{k}^{z}$ in Eq.\ (\ref{M_eq}) for the evaluation of the order parameter $M$, for example.  Such similarity-transformed operators may be expanded as the well-known nested commutator sums
\begin{equation}
{\rm e}^{-S}\Omega{\rm e}^{S} = \sum_{n=0}^{\infty}\frac{1}{n!}[\Omega,S]_{n}\,, \label{nest_commute_sums}
\end{equation}
where $[\Omega,S]_{n}$ is the $n$-fold nested commutator, defined iteratively as
\begin{equation}
[\Omega,S]_{n}=[[\Omega,S]_{n-1},S]\,; \quad [\Omega,S]_{0}=S\,.
\end{equation}
It is important to realize that in practice, for all usual choices of
the operator $\Omega$, the otherwise infinite sum in Eq.\
(\ref{nest_commute_sums}) will actually terminate {\it exactly} at a
(low) finite order.  The reasons for this are that the operator
$\Omega$ usually contains only finite-order multinomial terms in the
corresponding sets of single-spin operators (as for $H$ here), and
that all of the elements in the decomposition of $S$ in Eq.\
(\ref{ket_eigen_eq}) mutually commute.  The SU(2) commutation
relations for the spin operators then readily imply that the sum in
Eq.\ (\ref{nest_commute_sums}) will terminate after a finite number of
terms.

Thus, the {\it only} approximation that we need to make in order to
implement the CCM is to restrict the set of multispin-flip
configurations $\{I\}$ that we retain in the expansions of Eqs.\
(\ref{ket_eigen_eq}) and (\ref{bra_eigen_eq}) for the correlation
operators $S$ and $\tilde{S}$, respectively, to some manageable
subset.  A well-tested such hierarchical scheme, which we will adopt
here, is the so-called localized lattice-animal-based subsystem
(LSUB$n$) scheme.  It retains all such multispin configurations that,
at the $n$th level of approximation, describe clusters of spins
spanning a range of no more than $n$ contiguous sites.  In this sense
a set of lattice sites is said to be contiguous if every site in the
set is NN to at least one other in the set (in some specified geometry
that defines NN pairs).  Hence, the configurations retained in the
LSUB$n$ scheme are those defined on all possible (polyominos or)
lattice animals up to size $n$.  Obviously, as the truncation index
grows without bound (i.e., as $n \rightarrow \infty$) the corresponding
LSUB$\infty$ limit is the exact result.

The space- and point-group symmetries of the lattice and of the model
state $|\Phi\rangle$ under study, together with any relevant
conservation laws, are used to minimize the effective size of the
index set $\{I\}$ that is retained at each LSUB$n$ level.  For
example, for our present Heisenberg interactions contained in Eq.\
(\ref{H_eq}) and for the N\'{e}el model state (in zero-external
field), the total $z$-component of spin,
$s_{T}^{z}\equiv \sum_{k=1}^{N}s_{k}^{z}$, is a conserved quantity,
where global spin axes are assumed, and hence we retain only multispin
configurations for the GS N\'{e}el phase with $s_{T}^{z}=0$.  Even
after incorporating all such symmetries and conservation laws, the
number $N_{f}=N_{f}(n)$ of distinct, nonzero fundamental
configurations that are included at a given $n$th level of LSUB$n$
approximation grows rapidly (and typically, super-exponentially) as a
function of the truncation index $n$.  For the N\'{e}el GS of the
spin-$\frac{1}{2}$ honeycomb-lattice monolayer, for example, we have
$N_{f}(10)=6\,237$ and $N_{f}(12)=103\,097$.  For the corresponding
bilayer case we have $N_{f}(8)=2\,560$ and $N_{f}(10)=70\,118$.  By
contrast, the canted N\'{e}el state (i.e., in the presence of a
transverse magnetic field) has less symmetries and hence the number
$N_{f}(n)$ for the calculation of the susceptibility $\chi$ is
appreciably higher than the corresponding number for the calculation
of the magnetic order parameter $M$ at the same level $n$ of
approximation.  Thus, for the spin-$\frac{1}{2}$ honeycomb lattice
monolayer, for the case of the canted N\'{e}el state as our CCM model
state, we have $N_{f}(8)=3\,304$ and $N_{f}(10)=58\,337$, whereas the
for corresponding bilayer case we have $N_{f}(6)=1\,494$ and
$N_{f}(8)=43\,338$.

Clearly, the use of both massive parallelization and supercomputing
resources is required for the derivation and solution of such large
sets of CCM equations for the GS correlation coefficients
$\{{\cal S}_{I},\tilde{{\cal S}}_{I}\}$.  We also use a purpose-built and
customized computer-algebra package \cite{ccm_code} for the derivation
of the equations to be solved [i.e., Eqs.\ (\ref{ket_eq}) and
(\ref{bra_eq_alternative})].  Previous work
\cite{Bishop:2017_honeycomb_bilayer_J1J2J1perp} on the current model,
based on the N\'{e}el state as CCM model state, was able to perform
LSUB$n$ calculations for the order parameter $M$, for example, for
values $n \leq 10$.  Due to the substantially decreased symmetry of
the canted N\'{e}el state, by contrast we are now able to perform
calculations for the transverse magnetic susceptibility $\chi$ only
for values $n \leq 8$.

The last step, and sole approximation, is now to extrapolate our
LSUB$n$ sequences of approximants for the calculated physical
parameters to the (exact) $n \rightarrow \infty$ limit.  For example,
for systems that display a GS order-disorder QPT, a well-tested and
accurate extrapolation scheme for the magnetic order parameter $M$ of
Eq.\ (\ref{M_eq}) has been found to be given by (and see, Refs.\
\cite{DJJFarnell:2014_archimedeanLatt,Li:2012_honey_full,Bishop:2012_honeyJ1-J2,RFB:2013_hcomb_SDVBC,Bishop:2017_honeycomb_bilayer_J1J2J1perp,DJJF:2011_honeycomb,PHYLi:2012_honeycomb_J1neg,Bishop:2012_honey_circle-phase})
\begin{equation}
M(n) = \mu_{0}+\mu_{1}n^{-1/2}+\mu_{2}n^{-3/2}\,.   \label{M_extrapo_frustrated}
\end{equation}
This scheme, which is also appropriate for phases whose magnetic order
parameter $M$ is zero or small, yields the respective LSUB$\infty$
extrapolant $\mu_{0}$ for $M$.  By contrast, a scheme with a leading
exponent of -1 [rather than the value $-\frac{1}{2}$ for $M$ in Eq.\
(\ref{M_extrapo_frustrated})] has been found (and see, e.g., Refs.\
\cite{Bishop:2015_honey_low-E-param,Bishop:2016_honey_grtSpins,Darradi:2008_J1J2mod,Farnell:2009_Xcpty_ExtMagField,Gotze:2016_triang})
to give excellent results for the zero-field transverse (uniform)
magnetic susceptibility of Eq.\ (\ref{chi_eq}) with $h=0$,
\begin{equation}
\chi(n) = x_{0}+x_{1}n^{-1}+x_{2}n^{-2}\,.   \label{X_extrapo}
\end{equation}
This scheme thus leads to the LSUB$\infty$ extrapolant $x_{0}$ as our
value for $\chi$.

Clearly, for each of the extrapolation schemes such as those in
Eqs. (\ref {M_extrapo_frustrated}) and (\ref{X_extrapo}), each of which
involves three fitting parameters, it is preferable to use four or
more input data points (i.e., LSUB$n$ approximants with different
values of the truncation parameter $n$).  However, the LSUB2 result is
usually likely to be too far removed from the $n \rightarrow \infty$
limit to be useful in the fits, if it can be avoided.  Nevertheless,
as we have remarked above, it is computationally infeasible to perform
LSUB$n$ calculations of $\chi$ for the spin-$\frac{1}{2}$ honeycomb
bilayer for values $n > 8$.  For these reasons our preferred set of
fitting values are those with $n=\{4,6,8\}$.  However, in all cases we
have also performed separate fits using data sets with
$n=\{2,4,6,8\}$.  The differences in the extrapolated values are
generally extremely small.

\section{RESULTS}
\label{results_section}
\begin{figure*}[!t]
\centerline{
\mbox{
\hspace{0.1cm}\subfigure[]{\includegraphics[width=6.7cm]{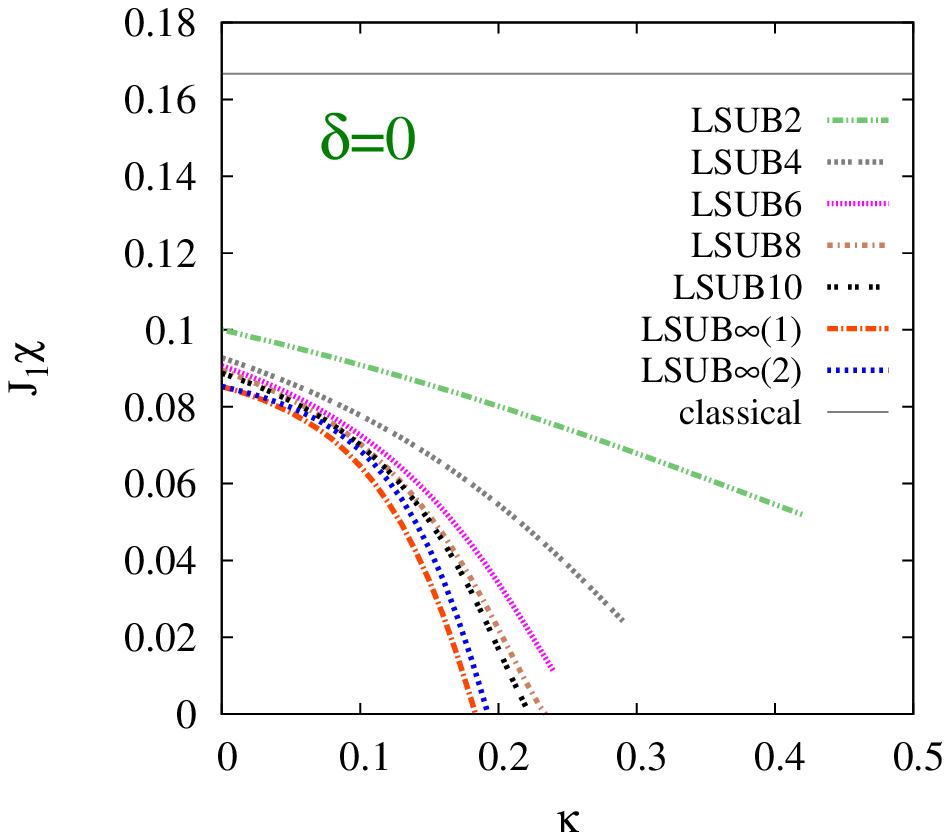}}
\hspace{-1.1cm}\subfigure[]{\includegraphics[width=6.7cm]{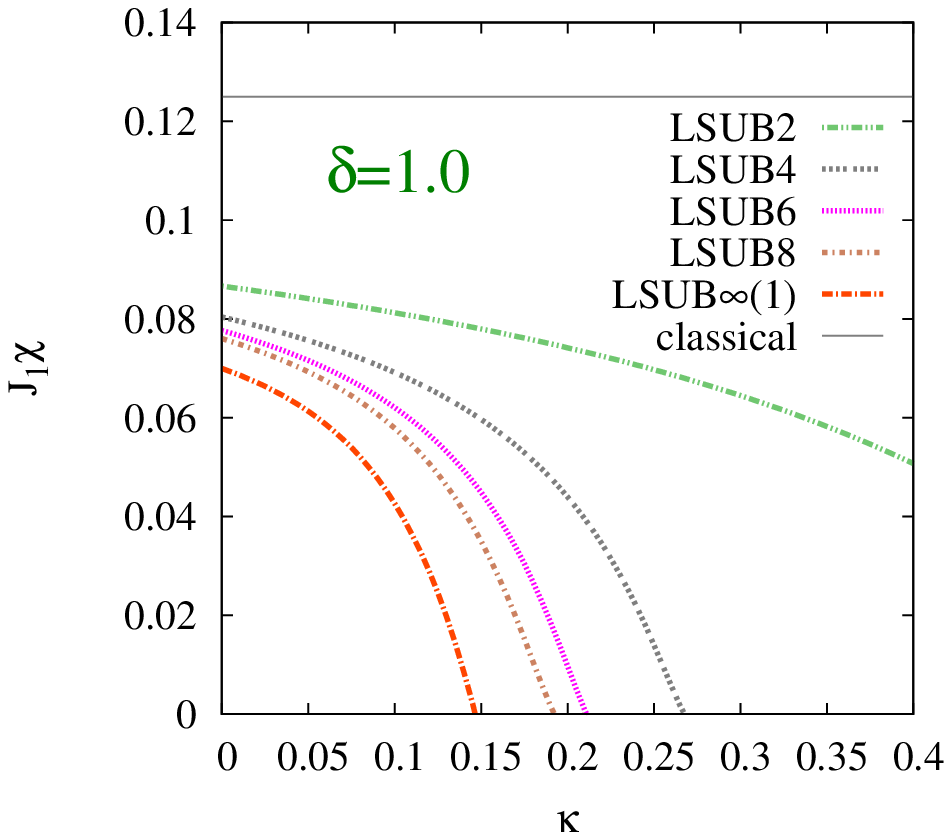}}
\hspace{-1.1cm}\subfigure[]{\includegraphics[width=6.7cm]{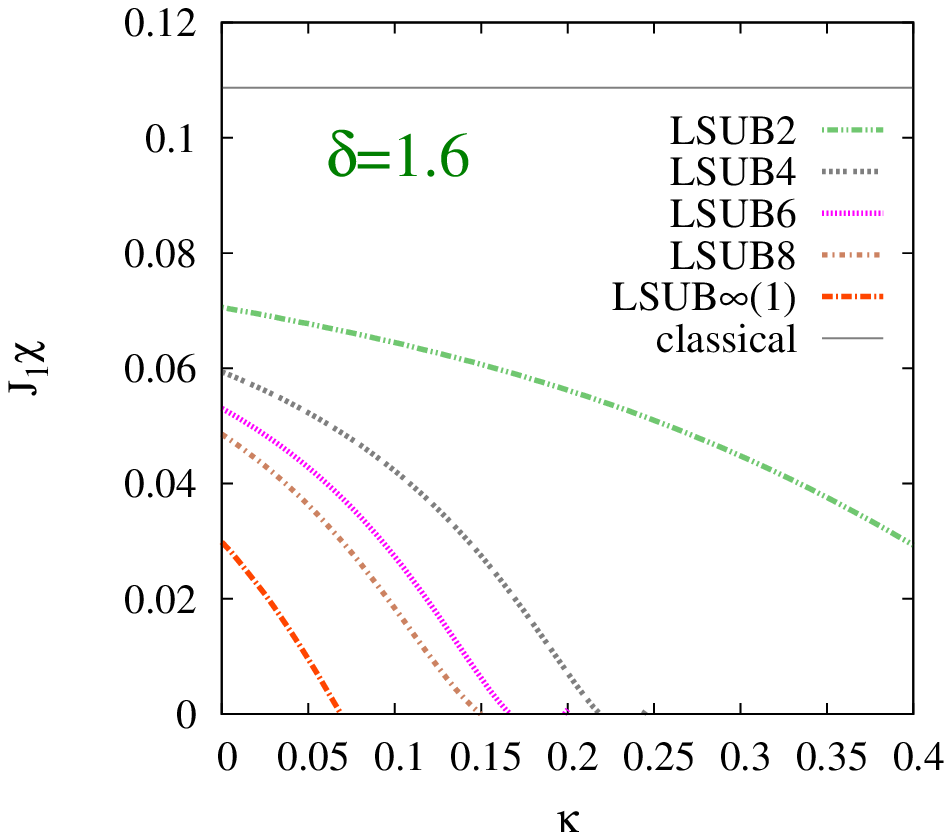}}
}
}
  \caption{CCM results for the zero-field transverse magnetic
  susceptibility $\chi$ (in units of $J_{1}^{-1}$) versus the frustration parameter $\kappa \equiv
  J_{2}/J_{1}$, for the spin-$\frac{1}{2}$
  $J_{1}$--$J_{2}$--$J_{1}^{\perp}$ model on the bilayer honeycomb
  lattice (with $J_{1}>0$), for three selected values of the scaled
  interlayer exchange coupling constant, $\delta \equiv J_{1}^{\perp}/J_{1}$:
  (a) $\delta=0$, (b) $\delta=1.0$, and (c) $\delta=1.6$.  Results
  based on the (canted) N\'{e}el state as CCM model state are shown in LSUB$n$
  approximations with $n=2,4,6,8$ (and also with $n=10$ for the
  special case of the $J_{1}$--$J_{2}$ monolayer, i.e., when
  $\delta=0$), together with various corresponding LSUB$\infty(i)$
  extrapolated results using Eq.\ (\ref{X_extrapo}) and the
  respective data sets $n=\{4,6,8\}$ for $i=1$ and $n=\{4,6,8,10\}$ for
  $i=2$ (for the case $\delta=0$ only).}
\label{chi_raw_extrapo_fix-J1perp}
\end{figure*}

We first show, in Fig.\ \ref{chi_raw_extrapo_fix-J1perp}, our results
for the zero-field transverse magnetic susceptibility $\chi$ as a
function of the intralayer frustration parameter $\kappa$, for three
respective values of the interlayer coupling parameter $\delta$.
In each case we also show the corresponding classical result from Eq.\
(\ref{chi_eq_neel}), which is now independent of $\kappa$ and thus
takes a constant value in each case.  For each of the three values of
$\delta$ shown in Figs.\ \ref{chi_raw_extrapo_fix-J1perp}(a),
\ref{chi_raw_extrapo_fix-J1perp}(b), and
\ref{chi_raw_extrapo_fix-J1perp}(c) we display our LSUB$n$ results for
values $n=2,4,6,8$, together with the LSUB$\infty(1)$ extrapolant
$x_{0}$ obtained from fitting Eq.\ (\ref{X_extrapo}) to the data set
$n=\{4,6,8\}$.  Uniquely, for the case $\delta=0$ shown in Fig.\
\ref{chi_raw_extrapo_fix-J1perp}(a), which corresponds to the
honeycomb-lattice monolayer, we are also able to perform calculations
at the LSUB10 level, which we also display there, together with a
separate LSUB$\infty(2)$ extrapolant $x_{0}$ obtained from fitting
Eq.\ (\ref{X_extrapo}) to the data set $n=\{4,6,8,10\}$.  Clearly, the
two extrapolations LSUB$\infty(1)$ and LSUB$\infty(2)$ are in
excellent agreement with each other.

Each of the cases shown in Fig.\ \ref{chi_raw_extrapo_fix-J1perp}, for
the three separate values of $\delta$, illustrates that the quantum
$(s=\frac{1}{2})$ values for $\chi$ are always substantially below the
corresponding classical $(s \rightarrow \infty)$ values.  More
striking, however, is that in each case there is a critical value
$\kappa_{c}(\delta)$ at which the extrapolated value for $\chi$
vanishes.  It is also clear from the case $\delta=0$ shown in Fig.\
\ref{chi_raw_extrapo_fix-J1perp}(a) that our extrapolations are quite
robust with respect to which LSUB$n$ data input sets are used.  The
vanishing of $\chi$, due to the strong effects of quantum
correlations, is, as we have noted in Sec.\ \ref{introd_sec}, a very
clear indication of the opening of a spin gap at this point
\cite{Mila:2000_M-Xcpty_spinGap,Bernu:2015_M-Xcpty_spinGap}, and we
may hence take it as an indicator of the QCP at which N\'{e}el order
melts.

\begin{figure*}[!t]
\centerline{
\mbox{
\hspace{0.1cm}\subfigure[]{\includegraphics[width=6.7cm]{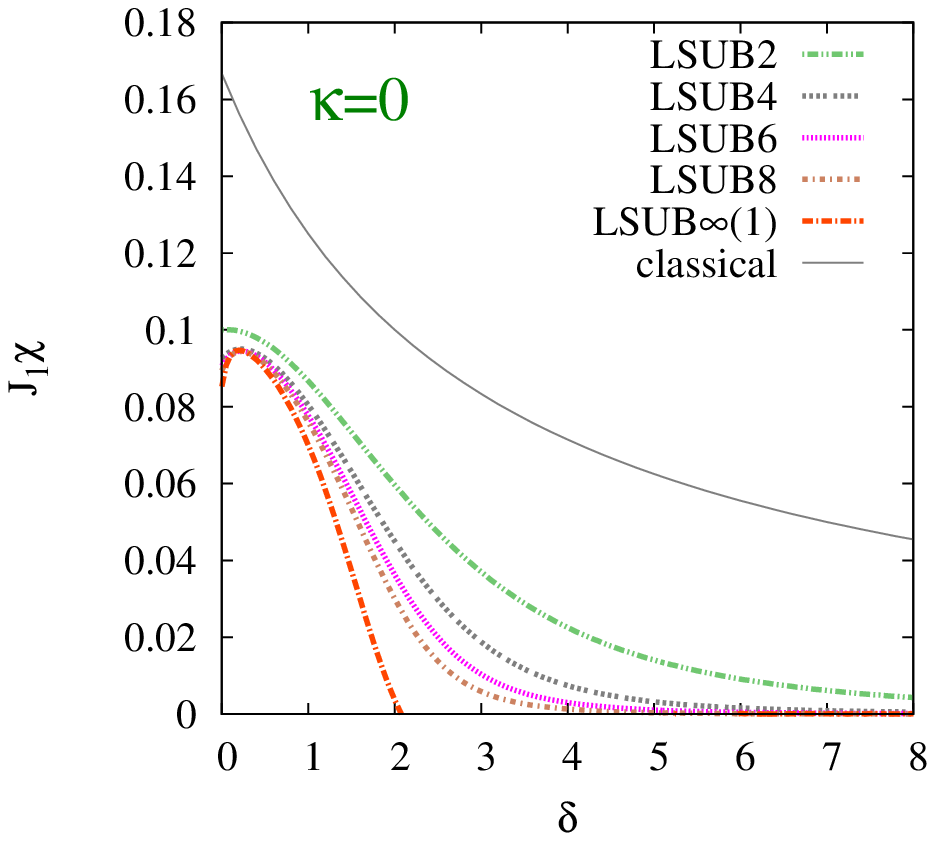}}
\hspace{-1.1cm}\subfigure[]{\includegraphics[width=6.7cm]{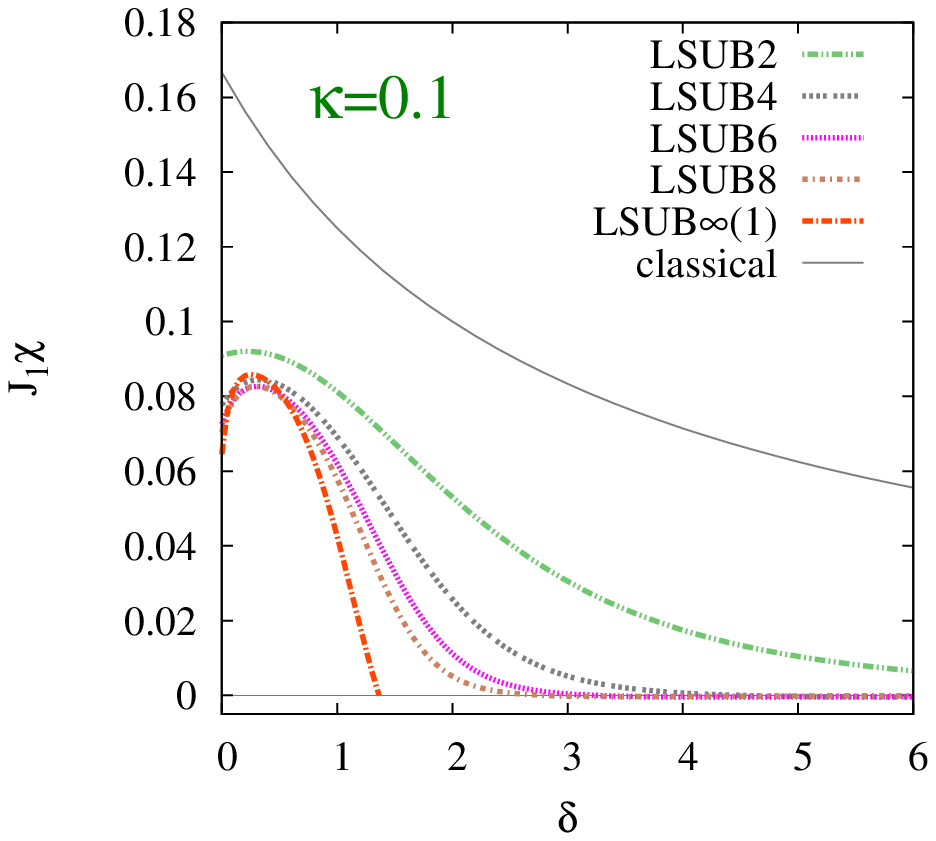}}
\hspace{-1.1cm}\subfigure[]{\includegraphics[width=6.7cm]{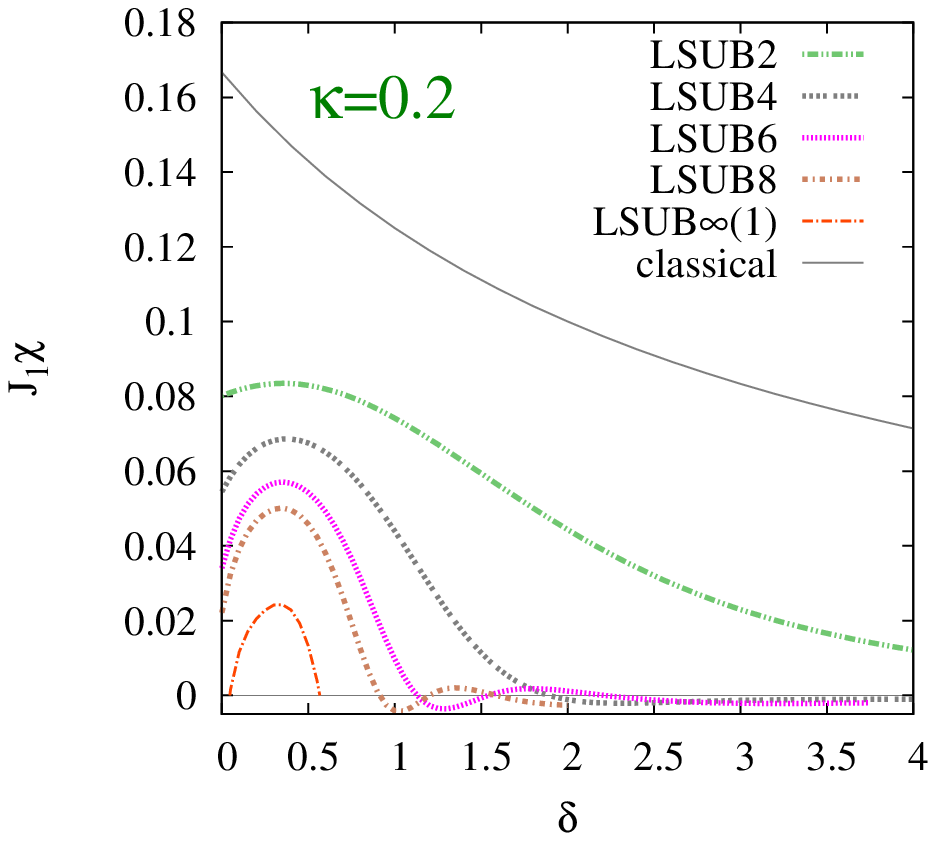}}
}
}
 \caption{CCM results for the zero-field transverse magnetic
  susceptibility $\chi$ (in units of $J_{1}^{-1}$) versus the scaled interlayer exchange coupling constant,
  $\delta \equiv J_{1}^{\perp}/J_{1}$, for the spin-$\frac{1}{2}$
  $J_{1}$--$J_{2}$--$J_{1}^{\perp}$ model on the bilayer honeycomb
  lattice (with $J_{1}>0$), for three selected values of the
  intralayer frustration parameter, $\kappa \equiv J_{2}/J_{1}$: (a)
  $\kappa=0$, (b) $\kappa=0.1$, and (c) $\kappa=0.2$.  Results
  based on the (canted) N\'{e}el state as CCM model state are shown in LSUB$n$
  approximations with $n=2,4,6,8$, together with corresponding LSUB$\infty(1)$
  extrapolated results using Eq.\ (\ref{X_extrapo}), with the
  data sets $n=\{4,6,8\}$.}
\label{chi_raw_extrapo_fix-J2}
\end{figure*}

The three cases shown in Fig.\ \ref{chi_raw_extrapo_fix-J1perp} reveal
that this critical value $\kappa_{c}(\delta)$ at which N\'{e}el order
vanishes decreases as the strength $\delta$ of the interlayer coupling
increases, at least for values of $\delta$ above a certain lower
critical value, to which we return in more detail below.  In Fig.\
\ref{chi_raw_extrapo_fix-J2} we show the effect of the interlayer
coupling separately, now for three illustrative values of the
intralayer frustration parameter $\kappa$.
In each case we show our CCM LSUB$n$ results with values $n=2,4,6,8$
of the truncation parameter, as well as the extrapolated value $x_{0}$
obtained from fitting the data points $n=\{4,6,8\}$ to Eq.\
(\ref{X_extrapo}), We also show the corresponding classical curves
obtained from Eq.\ (\ref{chi_eq_neel}).  Once again we see that the
effects of quantum correlations in the $s=\frac{1}{2}$ case are to
reduce the value of $\chi$ substantially from its classical
($s \rightarrow \infty$) value.

Figure \ref{chi_raw_extrapo_fix-J2}(a) shows our results for the case
$\kappa=0$ (i.e., without intralayer frustration), where NN AFM
interactions alone are present.  We observe the interesting feature
that as $\delta$ is slowly increased from zero the effect is first to
{\it increase} the value of $\chi$ both in absolute value and to bring
it closer to the corresponding classical value at the same value of
$\delta$.  This is presumably because the effects of quantum
correlations first weaken as $\delta$ is increased from zero, thereby
increasing the stability of N\'{e}el magnetic LRO.  This enhancement
reaches a maximum for each LSUB$n$ level of approximation (except for
the lowest-order, $n=2$) at a value $\delta \approx 0.5$.  N\'{e}el
LRO then reduces as $\delta$ is increased further.  At every LSUB$n$
level $\chi$ then tends asymptotically to zero.  It is evident that as
$n$ increases this asymptotic vanishing of $\chi$ becomes sharper and
sharper, ultimately as fully reflected in the LSUB$\infty(1)$
extrapolant that vanishes at the value
$\delta_{c}^{>}(\kappa=0) \approx 2.076$.  This agrees reasonably well
with a corresponding estimate $\delta_{c}^{>}(\kappa=0) \approx 1.645$
from a QMC calculation \cite{Ganesh:2011_honey_bilayer_PRB84}, which
can be performed only in the case of zero frustration ($\kappa=0$),
when the ``minus-sign problem'' is absent.

We may also compare our result for $\chi$ itself for the limiting case
$\kappa=0=\delta$ of a pure honeycomb-monolayer HAF with only NN
interactions.  Our extrapolated LSUB$\infty(1)$ result based on the
extrapolation scheme of Eq.\ (\ref{X_extrapo}) fitted to LSUB$n$ data
points with $n=\{4,6,8\}$ gives the value
$\chi(\kappa=0,\delta=0) \approx 0.0852/J_{1}$.  For this limiting
case alone we have also performed LSUB$n$ calculations with $n=10,12$
\cite{Bishop:2015_honey_low-E-param}.  For the LSUB12 calculation of
$\chi$ using the canted N\'{e}el state as CCM model state, for
example, the number of fundamental configurations is
$N_{f}(12)=1\,090\,448$.  The corresponding extrapolant using the
LSUB$n$ input data set with $n=\{6,8,10,12\}$ is
$\chi(\kappa=0,\delta=0)\approx 0.0847/J_{1}$, which again illustrates
the robustness of our results.  We are again in good agreement with a
corresponding result $\chi(\kappa=0,\delta=0)\approx 0.0778/J_{1}$
that was extracted (and see Ref.\ \cite{Bishop:2015_honey_low-E-param}
for details on how to do so) from a published QMC calculation of
L\"{o}w \cite{Low:2009_honey} for this unfrustrated case.

In Figs.\ \ref{chi_raw_extrapo_fix-J2}(b) and
\ref{chi_raw_extrapo_fix-J2}(c) comparable results to those in Fig.\
\ref{chi_raw_extrapo_fix-J2}(a) for the unfrustrated case ($\kappa=0$)
are also shown for the two cases when $\kappa=0.1$ and $\kappa=0.2$,
respectively.  As we would expect, as $\kappa$ is increased quantum
correlations become stronger and the susceptibility $\chi$ is reduced.
Correspondingly, the upper critical value $\delta_{c}^{>}(\kappa)$
above which a gapped state appears decreases monotonically with
increasing values of $\kappa$.  From the LSUB$\infty(1)$ extrapolation
in Fig.\ \ref{chi_raw_extrapo_fix-J1perp}(a) we see that
$\kappa_{c}(0) \approx 0.183$, which is below the value $\kappa=0.2$
shown in Fig.\ \ref{chi_raw_extrapo_fix-J2}(c).  What we now observe,
very interestingly, is that for values $\kappa > \kappa_{c}(0)$ that
are not too large, N\'{e}el order is re-established as $\delta$ is
increased above a lower critical $\delta_{c}^{<}(\kappa)$, while
remaining below the upper critical value $\delta_{c}^{>}(\kappa)$,
leading to the sort of reentrant behavior seen in Fig.\
\ref{chi_raw_extrapo_fix-J2}(c).  For the case $\kappa=0.2$ shown
there, for example, the LSUB$\infty(1)$ extrapolation gives the values
$\delta_{c}^{<}(0.2) \approx 0.046$ and
$\delta_{c}^{>}(0.2) \approx 0.567$.  Finally, as $\kappa$ is further
increased we arrive at an upper critical value $\kappa^{>}$ such that
$\delta_{c}^{<}(\kappa^{>}) = \delta_{c}^{>}(\kappa^{>})$, and for all
values $\kappa > \kappa^{>}$ a gapped paramagnetic state is present,
whatever the value of $\delta$, at least immediately beyond the
boundary of N\'{e}el stability.  Our LSUB$\infty(1)$ extrapolations
for $\chi$ lead to a value $\kappa^{>}\approx 0.212$, with
$\delta_{c}^{<}(\kappa^{>})=\delta_{c}^{>}(\kappa^{>})\approx 0.25(5)$.

\begin{figure}[!t]
  \includegraphics[width=12cm]{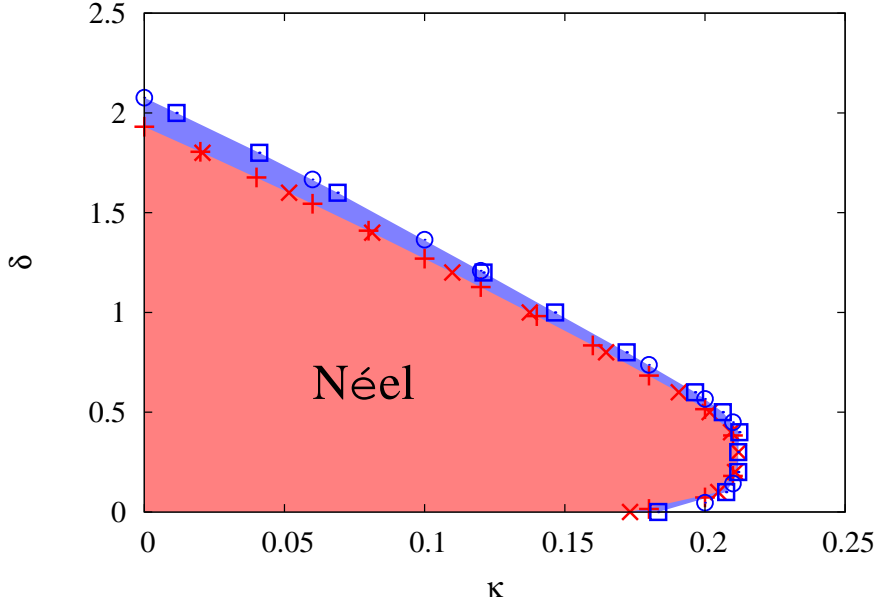}
  \caption{$T=0$ phase diagram of the spin-$\frac{1}{2}$
    $J_{1}$--$J_{2}$--$J_{1}^{\perp}$ model on the bilayer honeycomb
    lattice with $J_{1}>0$, $\delta \equiv J_{1}^{\perp}/J_{1}$, and
    $\kappa \equiv J_{2}/J_{1}$.   The blue open square ($\square$)
    symbols and the blue open circle plus ($\circ$) symbols are points at which the
    extrapolated  zero-field transverse magnetic susceptibility for the N\'{e}el
    phase vanishes, for specified values of $\delta$ and $\kappa$,
    respectively.  They thus represent the values $\kappa_{c}(\delta)$
    and $\delta_{c}^{>}(\kappa)$ [and also $\delta_{c}^{<}(\kappa)$
    for values of $\kappa$ in the range
    $\kappa_{c}(0) < \kappa < \kappa^{>}$], respectively.  In each
    case the (canted) N\'{e}el state is used as CCM model state, and Eq.\
    (\ref{X_extrapo}) is used for the extrapolations with
    the corresponding LSUB$n$ data sets $n=\{4,6,8\}$.  For
    comparison we also show by the red cross ($\times$)
    symbols and the red plus ($+$) symbols the points at which the
    extrapolated GS magnetic order parameter $M$ for the N\'{e}el
    phase vanishes, for specified values of $\delta$ and $\kappa$,
    respectively.  In each
    case the N\'{e}el state is used as CCM model state, and Eq.\
    (\ref{M_extrapo_frustrated}) is used for the extrapolations with
    the same LSUB$n$ data sets $n=\{4,6,8\}$.}
\label{phase_diag}
\end{figure}

Finally, in Fig.\ \ref{phase_diag}, we use our extrapolated
LSUB$\infty(1)$ results for $\chi$ such as those shown in Figs.\
\ref{chi_raw_extrapo_fix-J1perp} and \ref{chi_raw_extrapo_fix-J2} to
delineate the N\'{e}el phase boundary as the points where
$\chi \rightarrow 0$.
Different symbols are used to indicate the results for
$\kappa_{c}(\delta)$ at fixed values of $\delta$, as obtained from
curves such as those shown in Fig.\ \ref{chi_raw_extrapo_fix-J1perp},
and for both $\delta_{c}^{>}(\kappa)$ and $\delta_{c}^{<}(\kappa)$
(the latter in the case only when $\kappa_{c}(0)<\kappa<\kappa^{>}$),
as obtained from curves such as those shown in Fig.\
\ref{chi_raw_extrapo_fix-J2} for fixed values of $\kappa$.  The
overall accuracy of our results can be estimated from the fact that
points on the N\'{e}el phase boundary from two independent sets of
results agree so well with one another.  On Fig.\ \ref{phase_diag},
for comparison purposes, we also plot similar sets of points at which
the corresponding LSUB$\infty(1)$ extrapolants for the magnetic order
parameter $M$ [i.e., as determined from Eq.\ (\ref{M_extrapo_frustrated}) and LSUB$n$
data sets with $n=\{4,6,8\}$ used as input] vanish (and see Ref.\
\cite{Bishop:2017_honeycomb_bilayer_J1J2J1perp}).  It is extremely
gratifying that the N\'{e}el phase boundaries obtained from the points
where $\chi$ and $M$ vanish, respectively, are in such overall
excellent agreement.

\section{DISCUSSION AND SUMMARY}
\label{summary_section}
We have used the CCM and its well-defined and systematic LSUB$n$
hierarchy of approximations to investigate the N\'{e}el phase boundary
in the $T=0$ quantum phase diagram in the $\kappa\delta$ plane of the
spin-$\frac{1}{2}$ $J_{1}$--$J_{2}$--$J_{1}^{\perp}$ model on a bilayer
honeycomb lattice.  In particular, we have used the canted N\'{e}el
state (obtained from placing the N\'{e}el-ordered system in a
transverse external magnetic field) as our CCM model state in order to
calculate $\chi$, the transverse (uniform) magnetic susceptibility in
the zero-field limit.  Unlike in the classical
($s \rightarrow \infty$) version of the model, where $\chi$ never
vanishes, we find that for the $s=\frac{1}{2}$ model quantum
correlations become sufficiently strong to make $\chi \rightarrow 0$
along a curve in the $\kappa\delta$ plane.  All such points where
$\chi$ vanishes mark the emergence of a new gapped phase, and hence
the melting of N\'{e}el LRO.  We have exactly calculated
$\chi=\chi (\kappa,\delta)$ at high-order LSUB$n$ truncations with
$n \leq 8$, and as our sole approximation have extrapolated the
sequences of LSUB$n$ values for $\chi$ at given values of $\kappa$ and
$\delta$ with $n=\{4,6,8\}$, via a well-understood and well-tested
extrapolation scheme, to the limit $n \rightarrow \infty$ where the
method becomes exact in principle.  At all points along the N\'{e}el
phase boundary we have thereby seen that quasiclassical magnetic LRO
gives way to a nonclassical paramagnetic gapped state, which is almost
certainly a VBC state of one sort or another, and which almost
certainly differ as one moves along the boundary.  Thus, for the
large-$\delta$ region (for fixed values of $\kappa$) the N\'{e}el
state will certainly melt into a GS with IDVBC order, while for very small
values of $\delta$ it is most likely that the emergent gapped state
will have plaquette VBC (PVBC) order, as is generally agreed to be the
correct phase for the monolayer ($\delta=0$) for values of $\kappa$
beyond $\kappa_{c}(0)$ (and see, e.g., Refs.\ \cite{Bishop:2012_honeyJ1-J2,RFB:2013_hcomb_SDVBC}).

Perhaps the most striking feature of the phase diagram of Fig.\
\ref{phase_diag} is the marked reentrant behavior, whereby for values
of the intralayer frustration parameter in the range
$\kappa_{c}(0) < \kappa < \kappa^{>}$ there exists a range of values
of the interlayer coupling,
$\delta_{c}^{<}(\kappa)<\delta<\delta_{c}^{>}(\kappa)$ in which N\'{e}el LRO
is present.  Inside this region, which has larger values of
frustration present than the maximum allowed value $\kappa_{c}(0)$ for
N\'{e}el order in the monolayer, the effect of the bilayer coupling is
to enhance the N\'{e}el order to the extent that it reappears.  Beyond
a maximum value, $\kappa > \kappa^{>}$, however, no amount of
interlayer coupling suffices to re-establish N\'{e}el LRO.

We have also compared the N\'{e}el phase boundary that we have
obtained from the vanishing of $\chi$ with that obtained directly from
the vanishing of the N\'{e}el order parameter $M$.  In order to make a
valid comparison we have compared two completely independent sets of
CCM calculations for each quantity, both extrapolated with the same
sets of LSUB$n$ data with $n=\{4,6,8\}$ as input.  Figure \ref{phase_diag} shows the excellent level of agreement, which, in
turn, reinforces that at all points on the N\'{e}el phase boundary
shown, quasiclassical magnetic order gives way to a nonclassical
paramagnetic state with a nonzero energy gap to the lowest excited
state.  This is one of the most important findings of the present
study.

\section*{ACKNOWLEDGMENTS}
We thank the University of Minnesota Supercomputing Institute for the
grant of supercomputing facilities, on which the work reported here
was performed.  One of us (RFB) gratefully
acknowledges the Leverhulme Trust (United Kingdom) for the award of an
Emeritus Fellowship (EM-2015-007).  


\bibliographystyle{aipnum-cp}%
\bibliography{bib_general}

\end{document}